\documentclass[aps,prl,reprint,superscriptaddress]{revtex4-1}

\usepackage{hyperref}
\hypersetup{
    colorlinks,
    linkcolor={rgb:red,2; blue,1},
    citecolor={blue!60!black},
    urlcolor={blue!60!black}
}
\usepackage{graphicx}
\usepackage{amsmath,amssymb,amsthm,bbm,bm}
\usepackage{xcolor}
\usepackage{txfonts}
\usepackage{braket}  
\usepackage{dsfont}

\renewcommand{\H}{\hat{H}}

\newcommand{\ii}{\ensuremath{\mathrm{i}}}
\newcommand{\ee}{\ensuremath{\mathrm{e}}}
\newcommand{\dd}{\mathrm{d}}

\renewcommand{\Re}{\,\mathrm{Re}\,}
\renewcommand{\Im}{\,\mathrm{Im}\,}

\newcommand{\Tr}{\mathrm{Tr}}

\newcommand{\sg}{\ensuremath{\hat{\sigma}}}

\renewcommand{\H}{\ensuremath{\hat{H}}}
\renewcommand{\r}{\ensuremath{\hat{\rho}}}

\newcommand{\kk}{\ensuremath{\mathbf{k}}}

\newcommand{\rr}{\ensuremath{\mathbf{r}}}

\graphicspath{{./figs/}}

\begin{document}

\title{Taking the temperature of a pure quantum state}
\author{Mark T. Mitchison}
\email{mark.mitchison@tcd.ie}
\address{School of Physics, Trinity College Dublin, College Green, Dublin 2, Ireland}
\author{Archak Purkayastha}
\address{School of Physics, Trinity College Dublin, College Green, Dublin 2, Ireland}
\author{Marlon Brenes}
\address{School of Physics, Trinity College Dublin, College Green, Dublin 2, Ireland}
\address{Department of Physics and Centre for Quantum Information and Quantum Control, University of Toronto, 60 Saint George St., Toronto, Ontario, M5S 1A7, Canada}
\author{Alessandro Silva}
\affiliation{SISSA, Via Bonomea 265, I-34135 Trieste, Italy}
\author{John Goold}
\email{gooldj@tcd.ie}
\address{School of Physics, Trinity College Dublin, College Green, Dublin 2, Ireland}

\begin{abstract}
Temperature is a deceptively simple concept that still raises deep questions at the forefront of quantum physics research. The observation of thermalisation in completely isolated quantum systems, such as cold-atom quantum simulators, implies that a temperature can be assigned even to individual, pure quantum states. Here, we propose a scheme to measure the temperature of such pure states through quantum interference. Our proposal involves interferometry of an auxiliary qubit probe, which is prepared in a superposition state and subsequently decoheres due to weak coupling with a closed, thermalised many-body system. Using only a few basic assumptions about chaotic quantum systems --- namely, the eigenstate thermalisation hypothesis and the emergence of hydrodynamics at long times --- we show that the qubit undergoes pure exponential decoherence at a rate that depends on the temperature of its surroundings. We verify our predictions by numerical experiments on a quantum spin chain that thermalises after absorbing energy from a periodic drive. Our work provides a general method to measure the temperature of isolated, strongly interacting systems under minimal assumptions.
\end{abstract}

\maketitle

Advances in our understanding of thermodynamic concepts have always been inspired by the technologies of the time, from steam engines in the nineteenth century to ultra-cold atom simulators in the twenty-first. Irrespective of the historical era, the importance of measuring temperature cannot be overstated. In 1798, the American military man and scientist, Count Rumford, noticed that he could generate heat from friction while boring cannons in the arsenal of the Bavarian army he was tasked with reorganising. Rumford reported the systematic temperature increase of the water in which the cannon barrels were immersed~\cite{Thompson1798}, challenging the prevailing caloric theory of heat and inspiring James Joule to perform the decisive experiments that established energy conservation as the first law of a new thermodynamic theory. In his famous paddle-bucket experiment, Joule measured the mechanical equivalent of heat by observing the temperature change induced by stirring fluid in a thermally isolated container~\cite{Joule1850}. Here, we show that recasting Joule's experiment as a fully quantum-mechanical process leads to a fundamentally new scheme to measure the temperature of an isolated quantum many-body system. Our proposal relies on entangling the system with an auxiliary qubit that undergoes decoherence with a temperature-dependent rate. This thermometer scale is defined entirely through quantum interference and allows the measurement of temperature for generic systems in pure quantum states.

In the last two decades, experimental progress in cold-atom physics has enabled coherent quantum dynamics to persist over extraordinary timescales: long enough to observe isolated many-body systems thermalise without coupling to any external bath~\cite{Trotzky2012,Clos2016,Kaufman2016,Bordia2017,Tang2018}. The emergence of thermodynamics in this context is elegantly explained by the eigenstate thermalisation hypothesis (ETH)~\cite{Deutsch1991,Srednicki1994,Rigol2008}. The ETH posits that, in a sufficiently complex and chaotic system, each energy eigenstate encodes the properties of the equilibrium ensemble. As a result, local observables in a far-from-equilibrium scenario eventually thermalise under unitary evolution~\cite{DAlessio2016}. The final temperature is set by the energy density of the initial condition, which may be effectively a pure quantum state. Thermal fluctuations thus arise locally because of quantum entanglement between different parts of the system~\cite{Goldstein2006,Popescu2006} rather than by any classical statistical mixing. This begs the question: can the temperature of a pure state also be \textit{measured} in a completely quantum-mechanical way?

\begin{figure}[b]
    \centering
    \includegraphics[width=\linewidth]{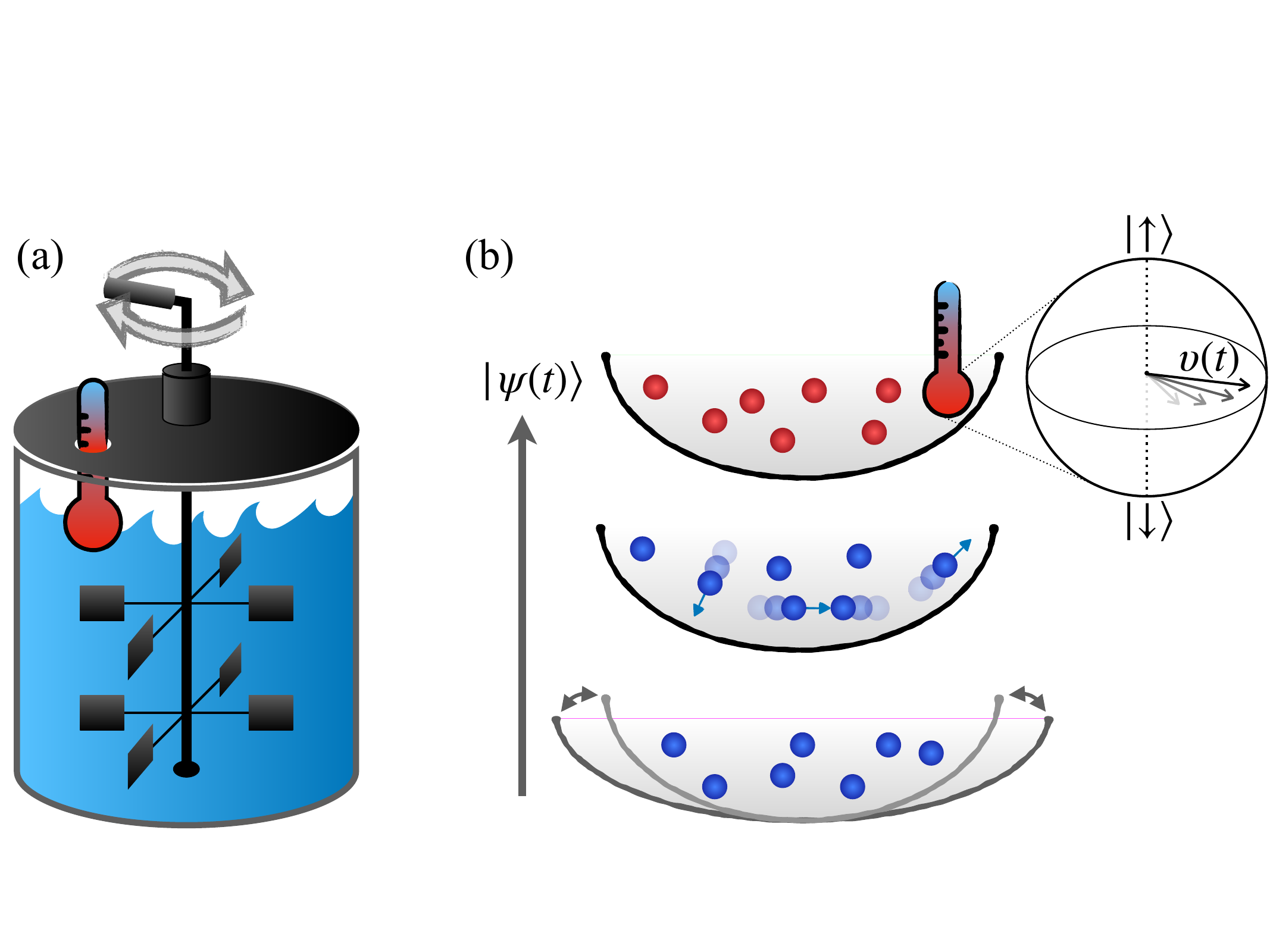}
\caption{Illustration of an experiment where work is performed on a thermally isolated system --- such as (a)~a bucket of water or (b)~an ultracold atomic gas --- thus driving it into a non-equilibrium state.  After the external force is removed, collisions between particles lead to irreversible thermalisation at a temperature determined by the energy density of the initial state, even though in (b) the global evolution is unitary and the system is described by a pure quantum state. The final temperature can be inferred by entangling the system to a qubit probe and measuring the resulting decoherence rate.}
    \label{fig:schematic}
\end{figure}

Our pure-state thermometry scheme, depicted in Fig.~\ref{fig:schematic}, draws inspiration from Joule's pioneering experiment, for which thermal isolation was vital. We consider the extreme case of an isolated quantum system such as an ultra-cold atomic gas. Work is performed by changing some external constraint, thus driving the system out of equilibrium in analogy to Joule's paddles. The driving force is then removed and the system relaxes under unitary evolution. Local observables thermalise to a temperature governed by the work performed, i.e.~the mechanical equivalent of heat. Joule's apparatus included an \textit{in situ} thermometer to measure the temperature change of the insulated fluid. In our setup, this role is played by an auxiliary qubit that becomes entangled with the many-body system. Assuming only the ETH and the equations of diffusive hydrodynamics, we show that the qubit undergoes pure exponential decoherence at a temperature-dependent rate that can be interferometrically measured~\cite{Cetina2015,Cetina2016,Skou2021}, providing a uniquely quantum thermometer for pure states.

Our work contributes to a growing body of literature seeking to establish the fundamental quantum limits of thermometry~\cite{Mehboudi2019}. The traditional approach --- used in Joule's measurements, for example --- is to let the thermometer exchange energy with its surroundings and wait for equilibration. Unfortunately, this becomes challenging to implement at low temperature, where a precise thermometer needs small energy scales and correspondingly long thermalisation times~\cite{Correa2015}. These drawbacks can be avoided by inferring temperature from the non-equilibrium dynamics of a probe, assuming a reliable model of the process is available~\cite{Bruderer_2006, Stace2010,Sabin2014,Hangleiter2015,Jevtic2015, Johnson2016,Razavian2019,Mitchison2020,Bouton2020,Adam2021}. In particular, Refs.~\cite{Johnson2016,Razavian2019,Mitchison2020} have shown that pure decoherence dynamics can encode temperature with a precision that is completely independent of the probe's energy. However, these proposals require the thermal system to be described by the canonical ensemble, as appropriate for an open system coupled to a heat reservoir. In contrast, our protocol offers a general solution to the problem of thermometry for isolated quantum systems, without the inherent limitations of small thermal probes that equilibrate with the system.

\begin{figure}[t!]
	\centering
	\includegraphics[width=\linewidth]{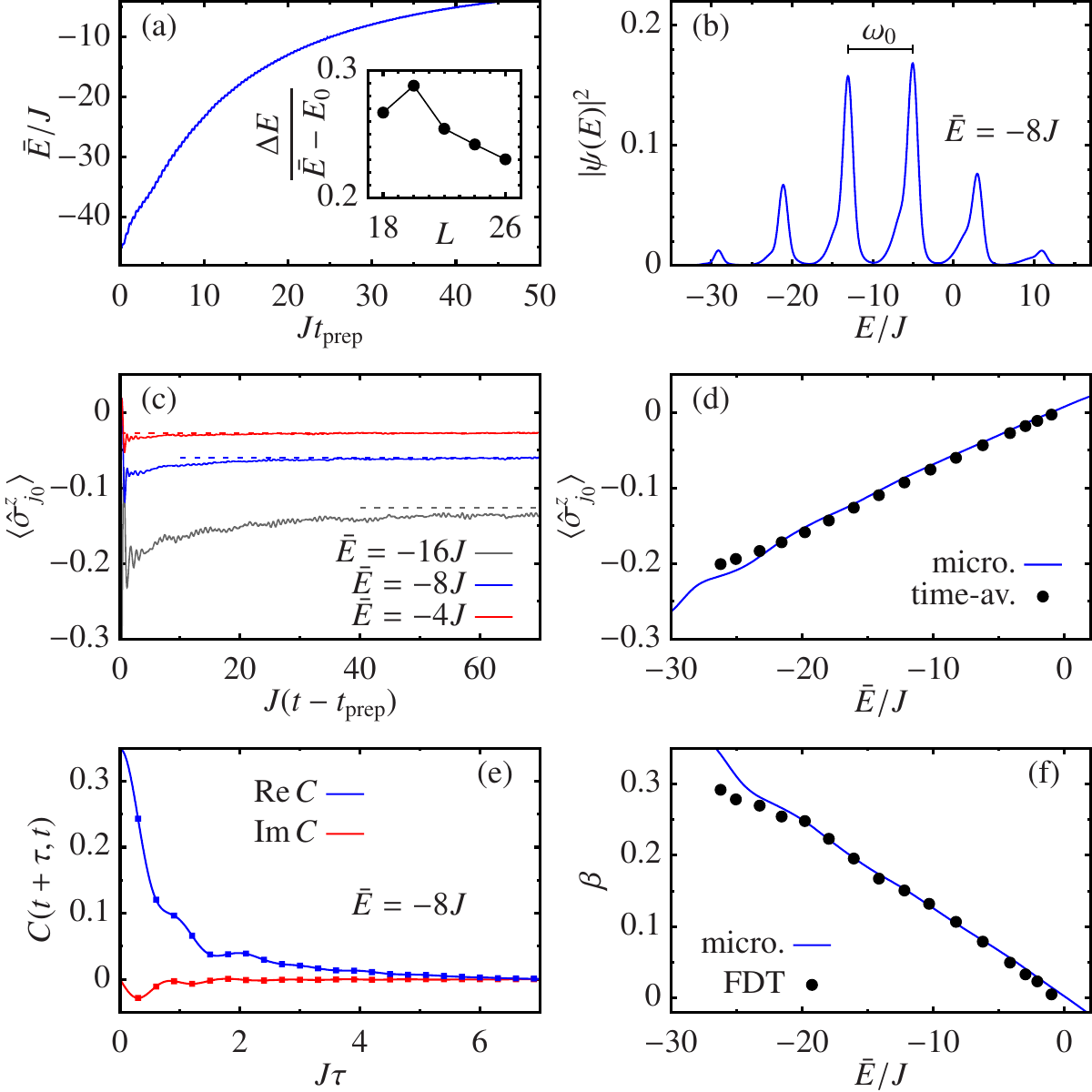}
	\caption{Unitary heating of a quantum spin-$\tfrac{1}{2}$ chain. (a)~Mean energy $\bar{E} =\braket{\psi(t_{\rm prep})|\hat{H}|\psi(t_{\rm prep})}$ of the chain as a function of the preparation time $t_{\rm prep}$ under local driving, $\hat{H}(t) = \hat{H} + a \sin(\omega_0 t)\sg_{j_0}^z$, applied to one site, $j_0$. Inset: Energy fluctuations, $\Delta E^2 = \braket{\psi(t_{\rm prep})|(\hat{H}-\bar{E})^2|\psi(t_{\rm prep})}$, versus system size at fixed temperature $T(\bar{E})=10J$. (b)~Energy distribution of the prepared state, $\nolinebreak{|\psi(E)|^2 = \sum_n |\braket{E_n|\psi(t_{\rm prep})}|^2\delta(E-E_n)}$, where $\hat{H}\ket{E_n} = E_n\ket{E_n}$. (c)~Equilibration of the local magnetisation after the drive is switched off. Solid lines show the dynamics of $\langle \hat{\sigma}^z_{j_0}\rangle$, with $\bar{E}$ increasing from the bottom to the top line. Dashed lines show the corresponding microcanonical average. (d)~Time-averaged local magnetisation after equilibration (black dots, obtained by time-averaging over an interval $\delta t \geq 20 J^{-1}$) compared with the microcanonical average (blue line). (e)~Auto-correlation function $C(t+\tau,t)$ of the local operator $\hat{A}=\sum_j u_j \sg^z_j$, where $u_j \propto \ee^{-(j-j_0)^2}$ is a Gaussian profile ($\sum_ju_j=1$). Lines show the real (blue/upper line) and imaginary (red/lower line) parts of $C(t+\tau,t)$ for $t- t_{\rm prep}=100J^{-1}$, while squares indicate near-identical values for $t- t_{\rm prep}=110J^{-1}$. (f)~Inverse temperature estimated by fitting the low-frequency noise and response functions to the FDT $\tilde{\chi}''(\omega)/\tilde{S}(\omega)= \tanh(\beta\omega/2)$ (black dots) and the corresponding microcanonical prediction (blue line). Parameters: $\Delta =0.55J$, $h=J$, $\omega_0=8J$, $a=2J$.}
	\label{fig:pure_thermo} 
\end{figure}

\textit{Spin-chain example.---}The quantum equivalent of Joule's paddle bucket is best illustrated by a specific example, although our scheme is general. Fig.~\ref{fig:pure_thermo} details an \textit{in silico} experiment where a thermally isolated many-body system is heated by periodic driving~\cite{Bunin2011,DAlessio2014,Lazarides2014}. We simulate an archetypal model of a quantum chaotic system: a Heisenberg spin-$\tfrac{1}{2}$ chain~\cite{Jepsen2020,Scheie2021} with Hamiltonian ($\hbar=k_B=1$)
\begin{equation}
\hat{H} =  J \sum_{j=1}^{L} \left (\sg_j^x \sg_{j+1}^x + \sg_{j}^y \sg_{j+1}^y + \Delta \sg_j^z \sg_{j+1}^z \right ) +  h \sum_{j\, \rm odd} \sg_j^z,
\end{equation}
where $\sg^{x,y,z}_j$ are Pauli operators pertaining to lattice site $j$. The exchange coupling $J$ and anisotropy $J\Delta$ respectively describe the kinetic and interaction energy of conserved spin excitations, while $h$ is a staggered magnetic field that breaks integrability~\cite{Brenes2020prl}. By exploiting Runge-Kutta methods for time evolution~\cite{Elsayed2013,Steinigeweg2014a,Steinigeweg2014b,Steinigeweg2015} and the kernel polynomial method to evaluate thermal and spectral properties~\cite{Weisse2006,Yang2020}, our simulations probe thermalisation dynamics at system sizes beyond those accessible to exact diagonalisation. Numerical methods are described in the Supplemental Material~\cite{SM}.

At time $t=0$, the chain is prepared in its ground state with energy $E_0$. An oscillatory field is then applied locally, pumping energy steadily into the system until the drive is switched off at time $t_{\rm prep}$ [Fig.~\ref{fig:pure_thermo}(a)]. This procedure generates a class of non-equilibrium pure states whose average energy $\bar{E}$ can be selected by tuning the preparation time. These states have a structured energy distribution featuring sharp peaks spaced by the drive frequency [Fig.~\ref{fig:pure_thermo}(b)]. Importantly, the corresponding energy fluctuations $\Delta E$ are sub-extensive, meaning that~$\Delta E/(\bar{E}-E_0)$ decreases with system size [Fig.~\ref{fig:pure_thermo}(a) inset].

After the drive is switched off, the system evolves autonomously and local observables relax to equilibrium [Fig.~\ref{fig:pure_thermo}(c)], exhibiting small fluctuations around a value that is close to the prediction of the microcanonical ensemble [Fig.~\ref{fig:pure_thermo}(d)]. This ensemble is characterised by a single parameter: the average energy, $\bar{E}$, with the corresponding inverse temperature $T^{-1}\equiv\beta = \beta(\bar{E})$ given by the fundamental definition $\beta(E) = \dd\mathcal{S}/\dd E$, where $\mathcal{S}(E)$ is the microcanonical entropy. Similar thermal behaviour is observed in correlation functions like $C(t',t) = \braket{ \hat{A}(t')\hat{A}(t)} - \braket{ \hat{A}(t')}\braket{\hat{A}(t)}$, with $\hat{A}$ a local observable, which become approximately stationary at long times, i.e. $C(t+\tau,t) \approx C(\tau)$ [Fig.~\ref{fig:pure_thermo}(e)]. Conventionally, one writes $C(\tau)$ in terms of the symmetrised noise function $S(\tau) = \Re[C(\tau)]$ and the dissipative response function $\chi''(\tau) = \ii \Im[C(\tau)]$. After relaxation, their Fourier transforms are related by the fluctation-dissipation theorem (FDT), $\tilde{S}(\omega) = \coth(\beta\omega/2) \tilde{\chi}''(\omega)$, as expected in thermal equilibrium [Fig.~\ref{fig:pure_thermo}(f)].

\begin{figure}
	\centering
	\includegraphics[width=\linewidth]{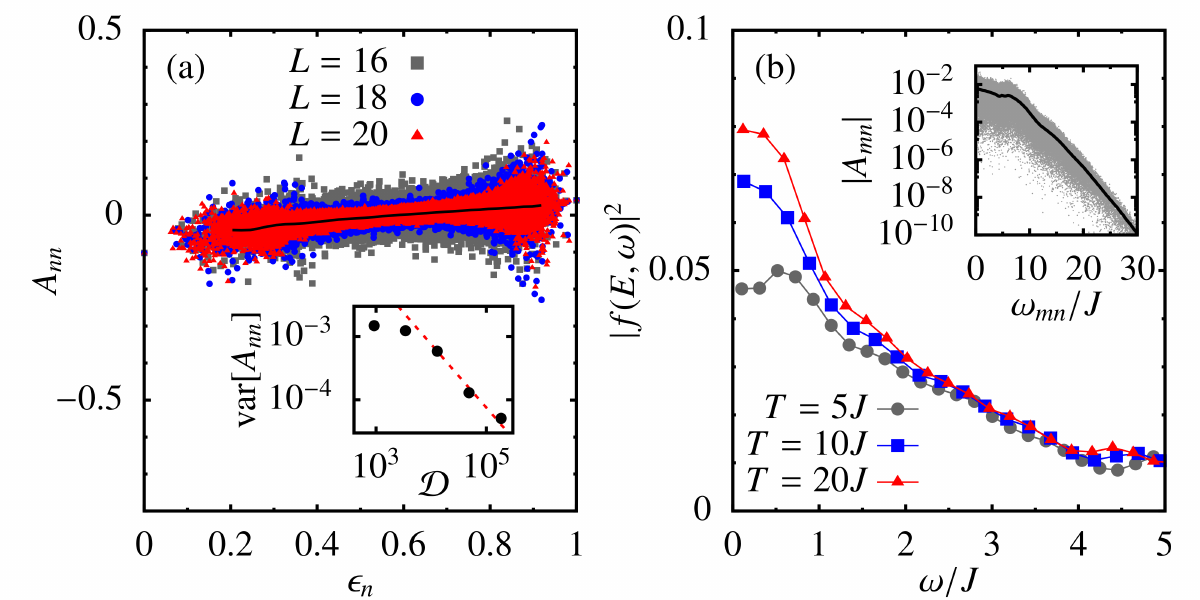}
	\caption{Eigenstate thermalisation in the staggered-field Heisenberg spin chain [see Fig.~\ref{fig:pure_thermo} caption for details]. (a)~Diagonal matrix elements of the local operator $\hat{A}$ concentrate around a smooth function (black line) of the energy density, $\epsilon_n = (E_n-E_{\rm min})/(E_{\rm max}-E_{\rm min})$. Inset: Variance of diagonal elements evaluated within the central 10\% of the spectrum for different system sizes, showing the scaling ${\rm var}[A_{nn}]\sim\mathcal{D}^{-1}$ (dashed red line). (b) Low-frequency spectral function for $L=18$ and three different temperatures. Inset: Off-diagonal elements near $T=5J$ (grey points; only 1\% of elements shown) and a running average of $|A_{mn}|$ (black line).}
	\label{fig:ETH}
\end{figure}

The thermalisation of these ``paddle-bucket'' preparations is striking in light of the highly non-equilibrium energy distribution displayed in Fig.~\ref{fig:pure_thermo}(b). Nevertheless, this behaviour is completely generic and fully explained by the ETH, which can be formulated as an ansatz for the matrix elements of an arbitrary local observable, $\hat{A}$, in the energy eigenbasis~\cite{Srednicki1999}, i.e.~$A_{mn}=\langle E_m|\hat{A}|E_n\rangle$, where $\hat{H}\ket{E_n} = E_n\ket{E_n}$. The ansatz reads as
\begin{equation}
\label{ETH}
A_{mn} = \begin{cases}\hspace{2mm}
A(E_{n}) + \mathcal{O}(\mathcal{D}^{-1/2}), & m=n, \\
\hspace{2mm}  \ee^{-\mathcal{S}(E_{mn})/2} f(E_{mn},\omega_{mn}) R_{mn} + \mathcal{O}(\mathcal{D}^{-1}), & m\neq n,
\end{cases}
\end{equation}
where $A(E_{n})$ and $f(E_{mn},\omega_{mn})$ are smooth functions of their arguments, $E_{mn} = \tfrac{1}{2}(E_m+E_n)$ and $\omega_{mn} = E_m-E_n$, while $R_{mn}$ is a Hermitian matrix of random numbers with zero mean and unit variance, and $\mathcal{D}$ is the Hilbert-space dimension. See Fig.~\ref{fig:ETH} for an example and Ref.~\cite{SM} for further details. As is well known~\cite{DAlessio2016}, the ETH~\eqref{ETH} implies that any highly excited state with sub-extensive energy fluctuations will thermalise under unitary dynamics. More precisely, the expectation value of a local observable converges to its time average $\overline{ \braket{\hat{A}}} = \sum_n |\braket{E_n|\psi}|^2 A_{nn} =  A(\bar{E}) + \mathcal{O}(\Delta E^2/\bar{E}^2_*)$, with $A(\bar{E})$ equal to the microcanonical average at inverse temperature $\beta(\bar{E})$, while the spectral function $f(\bar{E},\omega)$ determines the noise and response functions (up to sub-extensive corrections) as~\cite{DAlessio2016,Brenes2020prl}
\begin{align}
    \label{noise_spectrum}
    &\tilde{S}(\omega) = 2\pi \cosh(\beta\omega/2)|f(\bar{E},\omega)|^2, \\
    \label{response_function}
    &\tilde{\chi}''(\omega)  = 2\pi \sinh(\beta\omega/2)|f(\bar{E},\omega)|^2,
\end{align}
immediately implying the FDT. Although these features of the ETH have long been understood, the low-frequency behaviour of the spectral function has only recently been identified as a sensitive indicator of quantum many-body chaos~\cite{Brenes2020prb,Pandey2020}. For a generic observable in a non-integrable system, $f(E,0)$ is non-zero and may vary significantly with temperature [Fig.~\ref{fig:ETH}(b)]. This observation forms the basis of our thermometry scheme.

\textit{Thermometry protocol.---}Our thermometer comprises a qubit with energy eigenstates $\ket{\uparrow}$ and $\ket{\downarrow}$, coupled to the system by an interaction of the form $\hat{H}_{\rm int} = \ket{\uparrow}\bra{\uparrow} \otimes g\hat{A}$ for some local observable $\hat{A}$ and coupling constant $g$. This kind of interaction --- which can be engineered, for example, using Feshbach resonances in ultracold gases~\cite{Cetina2016} --- conserves the qubit's energy and ensures that it does not participate in the dynamics while in its ground state $\ket{\downarrow}$. Suppose that at time $t_0$, the thermal system of interest is in the pure state $\ket{\psi(t_0)}=\ket{\psi_0}$. The protocol begins by exciting the qubit into a superposition $\ket{+} = \tfrac{1}{\sqrt{2}}(\ket{\uparrow}+\ket{\downarrow})$ with a $\pi/2-$pulse, preparing the joint product state $\ket{\Psi(t_0)}= \ket{+}\ket{\psi_0}$. In a frame rotating at the qubit precession frequency, the Schr\"odinger evolution is then $\ket{\Psi(t)} = \frac{1}{\sqrt{2}}\left(\ee^{-\ii\H (t-t_0)}\ket{\downarrow}\ket{\psi_0} +\ee^{-\ii(\H+g\hat{A}) (t-t_0)}\ket{\uparrow}\ket{\psi_0}  \right)$. Entanglement develops between the probe and the system, leading to a loss of distinguishability quantified by the fidelity between many-body system states
\begin{equation}
\label{fidelity}
|\varv(t)|^2 = |\braket{\psi_0|\ee^{\ii\H (t-t_0)}\ee^{-\ii(\H+g\hat{A}) (t-t_0)}|\psi_0}|^2.
\end{equation}
The resulting decrease in interference contrast is reflected in the off-diagonal elements of the qubit density matrix, $\hat{\rho}_{\rm q}(t)=\Tr_{\rm sys}\,\ket{\Psi(t)}\bra{\Psi(t)}$, which decay in time according to $\braket{\downarrow\rvert\r_{\rm q}(t)\lvert\uparrow} = \tfrac{1}{2}\varv(t)$. This decoherence is finally probed by applying a second $\pi/2$-pulse with a phase $\theta$ relative to the first one, then measuring the excited-state probability of the qubit, $P_\uparrow = \tfrac{1}{2}(1+\Re[\ee^{\ii \theta}\varv(t)]).$ The time-dependent overlap $\varv(t)$ is thus reconstructed by varying $\theta$.

\begin{figure}
	\centering
	\includegraphics[width=\linewidth]{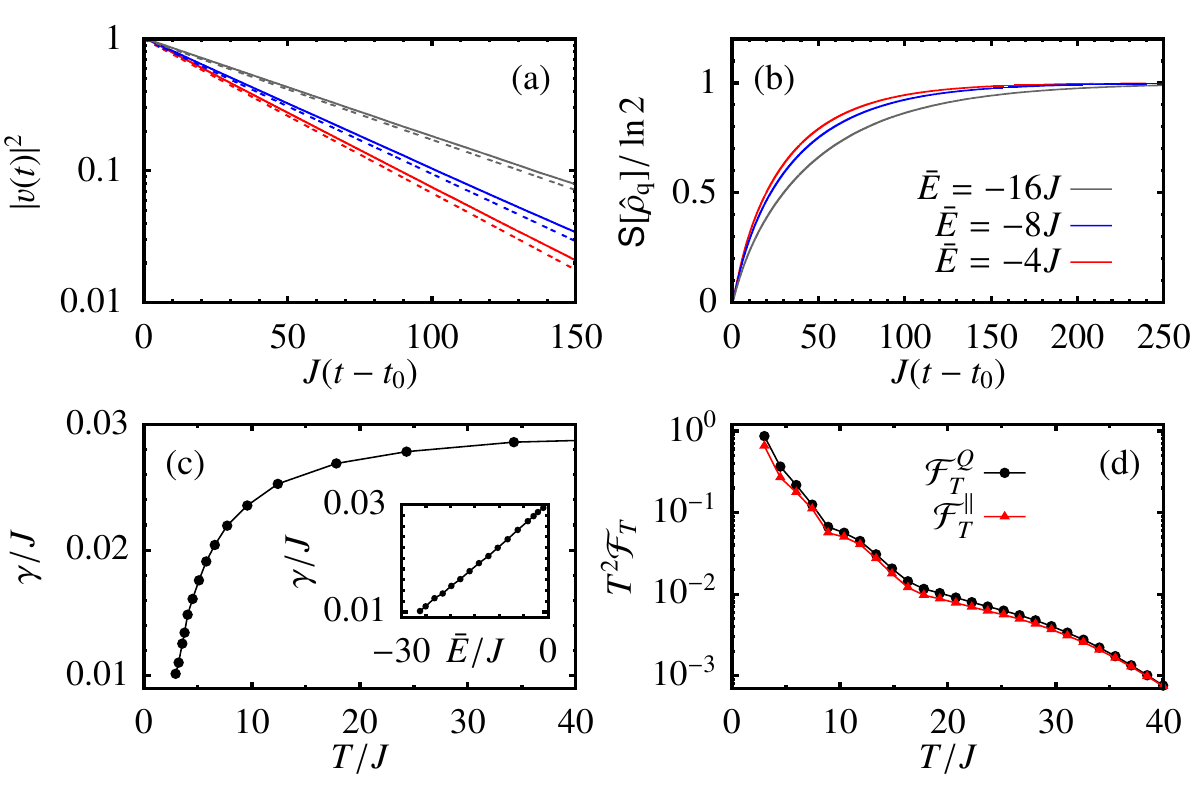}
	\caption{Decoherence of a qubit with coupling $g=0.2J$ to a spin-chain environment prepared in a pure thermal state as shown in Fig.~\ref{fig:pure_thermo}. (a)~The fidelity and (b)~the entanglement entropy for three different temperatures. Solid lines show an exact calculation of Eq.~\eqref{fidelity}, while the dashed lines show the weak-coupling approximation $|\varv(t)|^2 = \ee^{-\gamma (t-t_0)}$, with $\gamma=  g^2\tilde{S}(0)$ extracted from dynamical calculations of $C(\tau)$. We take $t_0-t_{\rm prep} = 100J^{-1}$ for $\bar{E}=-4J,-8J$ and $t_0-t_{\rm prep} = 200J^{-1}$ for $\bar{E}=-16J$. In (a), $\bar{E}$ increases from the top to the bottom line and vice versa in (b). (c)~Temperature dependence of the asymptotic decoherence rate, $\gamma=  g^2\tilde{S}(0)$, with the energy dependence as an inset. (d)~QFI (black dots) and Fisher information for a measurement in the qubit eigenbasis (red triangles), computed within the weak-coupling approximation as a function of temperature. Both quantities are evaluated at the time $t^*$ where the QFI is maximised, i.e. $\mathcal{F}^Q_T = \mathcal{F}^Q_T(t^*) \equiv \max_t\mathcal{F}^Q_T(t)$ and $\mathcal{F}^\parallel_T = \mathcal{F}^\parallel_T(t^*)$. }
	\label{fig:decoherence}
\end{figure}

\textit{Precision at weak coupling.---}To assess the temperature dependence of the interference contrast, we focus on the weak-coupling regime and approximate the fidelity~\eqref{fidelity} by a cumulant expansion to second order in $g$~\cite{SM}. We obtain $|\varv(t)|^2 = \ee^{-\Gamma(t)}$, where
\begin{align}
     \label{Gamma_def}
    & \Gamma(t) = 4g^2\int\frac{\dd \omega}{2\pi}\,\tilde{S}(\omega) \frac{\sin^2[\omega (t-t_0)/2]}{\omega^2}.
\end{align}
At weak coupling, the largest effects are seen for $t-t_0\gg \tau_c$, where $\tau_c$ is the characteristic timescale for the correlation function $C(\tau)$ to decay to zero. The integral in Eq.~\eqref{Gamma_def} is then dominated by the contribution near $\omega=0$, which implies pure exponential decoherence, $|\varv(t)|^2 \sim \ee^{-\gamma (t-t_0)}$, with an asymptotic decay rate $\gamma = g^2 \tilde{S}(0) \propto |f(\bar{E},0)|^2$. We numerically confirm this behaviour in Fig.~\ref{fig:decoherence}(a), which shows the fidelity for a probe coupled to a spin chain heated by the procedure of Fig.~\ref{fig:pure_thermo}. Even for moderate coupling strengths, we observe near-perfect exponential decay with a temperature-dependent rate in close agreement with the weak-coupling prediction. The decoherence is associated with a growth in the entanglement entropy $\mathsf{S}[\hat{\rho}_{\rm q}] = -\Tr[\hat{\rho}_{\rm q}\ln \hat{\rho}_{\rm q}]$, which saturates to the temperature-independent value $\mathsf{S}[\hat{\rho}_{\rm q}]\to\ln 2$ characterising a maximally entangled state [Fig.~\ref{fig:decoherence}(b)]. This distinguishes our non-equilibrium protocol from a thermalisation process. In Fig.~\ref{fig:decoherence}(c), the temperature dependence of the decoherence rate is analysed in more detail. We find that $\gamma$ depends almost linearly on energy density [Fig.~\ref{fig:decoherence}(c) inset], which translates into a non-linear variation with temperature [Fig.~\ref{fig:decoherence}(c) main panel] that is greatest at low temperatures.

We quantify the temperature information that can be extracted from our protocol using the quantum Fisher information (QFI). Consider a temperature estimate constructed from $M$ independent measurements in a given basis, $\mu$, on identical qubit preparations. For large $M$, the statistical error of any unbiased estimate is asymptotically bounded by $\Delta T^2 \geq 1/M\mathcal{F}^\mu_T \geq 1/M\mathcal{F}_T^Q$. Here, $\mathcal{F}^\mu_T$ is the Fisher information for the chosen basis while the QFI, $\mathcal{F}_T^Q = \max_\mu \mathcal{F}_T^\mu$, is the maximum over all measurements and thus describes the ultimate uncertainty limit imposed by quantum mechanics~\cite{Braunstein1994}. The temperature can be inferred from the exponential decay of $|\varv(t)|$ by measuring in the eigenbasis of $\hat{\rho}_{\rm q}(t)$, i.e.~by applying a final $\pi/2$-pulse with phase $\theta=-\arg\varv(t)$~\cite{SM}. Fig.~\ref{fig:decoherence}(d) shows the corresponding Fisher information, $\mathcal{F}^\parallel_T$, in the weak-coupling limit. Since $\mathcal{F}_T^\parallel\approx \mathcal{F}^Q_T$, we conclude that the decoherence rate captures almost all temperature information available from the probe in this example. For instance, we obtain the value $T^2\mathcal{F}^\parallel_T \approx  0.2$ at temperature $T=5J$, implying that $M= 500$ measurements could suffice to achieve a precision of $\Delta T/T \gtrsim 10\%$. Note that a single ultra-cold gas sample may host thousands of independent impurities~\cite{Cetina2016}. We emphasise that the achievable precision is independent of the qubit's energy gap, unlike a thermalised probe whose QFI depends exponentially on this gap at low temperature~\cite{Correa2015}.

\textit{Hydrodynamic decoherence.---}Our results show that the temperature of an isolated system can be measured using the most primitive features of quantum dynamics: namely, unitarily evolving wavefunctions and entanglement between subsystems. The scale of our thermometer is defined not through the energetic fluctuations of some statistical mixture, but by the rate of entanglement growth in a quantum decoherence process~\footnote{Note that here we refer to entanglement between the many-body system and the probe. This is a distinct concept from the entanglement entropy between subsystems within the many-body system, which also encodes temperature in a different sense~\cite{Garrison2018,Nakagawa2018} and can be interferometrically measured~\cite{Abanin2012,Pichler2016}.}. While this rate should generally increase with temperature, the precise dependence is system- and observable-specific. Nevertheless, since a generic system should display hydrodynamic behaviour at long times~\cite{Forster2018}, we can obtain a general form for $\gamma(T)$ assuming that the probe couples to diffusive modes of a conserved density. In $d=3$ spatial dimensions, we obtain~\cite{SM}
\begin{equation}
	\label{gamma_T}
	\gamma = \frac{2\bar{g}^2\chi_0 T}{D},
\end{equation}
where $D$ is the diffusion coefficient, $\chi_0$ is the thermodynamic susceptibility to long-wavelength density perturbations and $\bar{g}$ is a renormalised coupling that depends only on the probe's spatial profile. According to Eq.~\eqref{gamma_T}, the qubit's decoherence rate provides an ideal, linear thermometer scale within any temperature range where $D$ and $\chi_0$ are approximately constant, and allows for accurate thermometry in general whenever $D$ and $\chi_0$ are known as a function of temperature.

In low-dimensional systems --- such as our spin-chain example --- similar hydrodynamic arguments predict non-exponential decoherence at intermediate times, $\Gamma(t) \sim t^{3/2}$ for $d=1$ and $\Gamma(t) \sim t\ln t$ for $d=2$, which crosses over to pure exponential decay, $\Gamma(t)\sim \gamma t$, when $t\gtrsim \tau_c$~\cite{SM}. The asymptotic decoherence rate $\gamma$ depends on temperature as in Eq.~\eqref{gamma_T}, but both $\gamma$ and $\tau_c$ grow with the system size for $d<3$~\cite{SM}. However, $\tau_c$ is too small to clearly distinguish the crossover at system sizes accessible in our simulations, where only the long-time exponential decay is observed. This interesting competition of timescales calls for further research to characterise how Markovian dynamics~\cite{Srednicki1999,Nation2019pre,ParraMurillo2021} and thermodynamics~\cite{Iyoda2017,RieraCampeny2021} emerge for open quantum systems in chaotic environments.

\textit{Conclusion.---}Accurate, \textit{in situ} thermometry of isolated quantum systems is an outstanding problem in cold-atom physics, where strong, short-ranged correlations confound destructive global measurement techniques such as time-of-flight imaging. Conversely, a small quantum probe facilitates local, minimally destructive temperature measurements, in principle~\cite{Hohmann2016,Bouton2020}. Our proposal to infer temperature from decoherence dynamics does not require thermalisation of the qubit nor fine-tuning of its energy levels, and is applicable to generic many-body systems in arbitrary states with sub-extensive energy fluctuations. This opens a pathway for the toolbox of quantum-enhanced thermometry~\cite{Mehboudi2019} to probe the ultimate limit of an isolated system in a pure quantum state.

\textbf{Acknowledgements.} We thank S.~R.~Clark, C.~Jarzynski, A.~Polkovnikov, and J.~Richter for useful feedback on the manuscript. M.~B. and J.~G. thank M.~Rigol for stimulating their interest in the ETH. A.~P. acknowledges funding from the European Union's Horizon 2020 research and innovation programme under the Marie Sk{\l}odowska-Curie grant agreement No.~890884. Some calculations were performed on the Lonsdale cluster maintained by the Trinity Centre for High Performance Computing. This cluster was funded through grants from Science Foundation Ireland (SFI). We acknowledge the DJEI/DES/SFI/HEA Irish Centre for High-End Computing (ICHEC) for the provision of computational facilities, Project No.~TCPHY138A.  This work was supported by a SFI-Royal Society University Research Fellowship (J. G.) and the Royal Society (M. B.). J.~G. and M.~T.~M. acknowledge funding from the European Research Council Starting Grant ODYSSEY (Grant Agreement No. 758403) and the EPSRC-SFI joint project QuamNESS.

\bibliographystyle{apsrev4-1}
\bibliography{references}

\setcounter{secnumdepth}{3}
\setcounter{equation}{0}
\setcounter{figure}{0}
\section*{Supplemental Material}

\renewcommand{\theequation}{S\arabic{equation}}
\renewcommand{\thesubsection}{S\arabic{subsection}}
\renewcommand{\thefigure}{S\arabic{figure}}

\subsection{Numerical methods}
\label{app:numerics}

In this section we provide further details on the model and the methods used to analyse the quantitative examples discussed in the main text.

\subsubsection{Hamiltonian and observables}

The examples in the main text are based on the Hamiltonian
\begin{equation}\label{H_SF_app}
	\hat{H} =  J \sum_{j=1}^{L} \left (\sg_j^x \sg_{j+1}^x + \sg_{j}^y \sg_{j+1}^y + \Delta \sg_j^z \sg_{j+1}^z \right ) +  h \sum_{j\, \rm odd} \sg_j^z,
\end{equation}
with periodic boundary conditions. This Hamiltonian conserves the number of spin excitations, $\hat{N} = \tfrac{1}{2}\sum_j (1+\sg^z_j)$, and we work in the half-filled symmetry sector of states with eigenvalue $N = L/2$. The bulk parameters are chosen to be $h=J$ and $\Delta = 0.55$, for which the model is robustly non-integrable~\cite{Brenes2020prl}. Some of our examples focus on the local operator $\hat{A} = \sum_j u_j \sg_j^z$, where $u_j\propto \ee^{-(j-j_0)^2}$ is a Gaussian profile centred on site $j_0$, where $j_0 = L/2$ if $L/2$ is odd and $j_0=L/2+1$ if $L/2$ is even. To improve numerical efficiency, we set $u_j=0$ on all sites where $\ee^{-(j-j_0)^2} <10^{-3}$, and then normalise as $\sum_j u_j=1$. This generates an observable with support restricted to five sites of the lattice.

It is important to remark that, in order to obtain the function $f(E, \omega)$ from the coarse-grained average of the off-diagonal matrix elements of a local operator~\cite{Mondaini2017,Khatami2013,Brenes2020prl,Brenes2020prl2,Brenes2020prb}, all symmetries of the model should be resolved. Resolving these symmetries amounts to restricting the block-diagonal Hamiltonian to a single symmetry sub-sector of states corresponding to a given eigenvalue of the symmetry generator (performing a separate calculation for each block, if more than one is required). If the operators admix the symmetry sub-sectors~\cite{Leblond2020}, resolving the corresponding symmetry is not required to obtain $f(E, \omega)$ from the coarse-grained average of the off-diagonal matrix elements.

The function $f(E, \omega)$ obtained from this procedure is shown in Fig.~\ref{fig:ETH}(b). In contrast with the open-boundary chain, the model with periodic boundary conditions is translation-invariant. Instead of resolving this symmetry, we break it by augmenting the magnetic field acting on site $j=1$ by a small amount $\delta h = 0.1h$. Even with the addition of this small perturbation, in the zero-magnetisation sector, an underlying spatial reflection symmetry remains. This symmetry is broken by the operator $\hat{A}$ when $L / 2$ is odd, as in Fig.~\ref{fig:ETH}(b).

\subsubsection{Eigenstate thermalisation}
\label{app:ETH}

The ETH posits that the matrix elements of an observable $\hat{A}$ in the energy eigenbasis are of the form
\begin{equation}
\label{ETH-app}
A_{mn} = \begin{cases}\hspace{2mm}
A(E_{n}) + \mathcal{O}(\mathcal{D}^{-1/2}), & m=n, \\
\hspace{2mm}  \ee^{-\mathcal{S}(E_{mn})/2} f(E_{mn},\omega_{mn}) R_{mn} + \mathcal{O}(\mathcal{D}^{-1}), & m\neq n.
\end{cases}
\end{equation}
The meaning of each term on the right-hand side is illustrated in Fig.~\ref{fig:ETH} of the main text and explained as follows. The diagonal matrix elements ($m=n$) are given by a smooth function of energy, $A(E)$, up to fluctuations that scale inversely with the square root of the Hilbert-space dimension, $\mathcal{D}$ [Fig.~\ref{fig:ETH}(a)]. Therefore, all energy eigenstates near a given energy $E$ yield the same expectation value $\langle \hat{A}\rangle = A(E)$ in the thermodynamic limit. This identifies $A(E)$ as the microcanonical average of $\hat{A}$ at inverse temperature $\beta(E)$. In Fig.~\ref{fig:ETH}(a), this is indicated by the black line, which shows a running average of the diagonal matrix elements $A_{nn}$ within microcanonical windows of width $\delta \epsilon =0.02$ for a system size of $L=20$.

Meanwhile, the off-diagonal matrix elements ($m\neq n$) are exponentially small and erratically distributed [Fig.~\ref{fig:ETH}(b) inset], as described in Eq.~\eqref{ETH-app} by a Hermitian matrix $R_{mn}$ of random numbers with zero mean and unit variance. Underlying this distribution is a smooth spectral function $f(E,\omega)$ of the mean energy, $E_{mn} = \tfrac{1}{2}(E_m+E_n)$ and transition frequency, $\omega_{mn} = E_m-E_n$, which is revealed in the variance of the matrix elements within small energy and frequency windows [Fig.~\ref{fig:ETH}(b) main panel]. Specifically, the data in Fig.~\ref{fig:ETH}(b) are generated by finding all off-diagonal matrix elements at energy $E_{mn}$ consistent with a given temperature, $T=[\beta(E_{mn})]^{-1}$, then computing the variance, ${\rm var}[A_{mn}] = \ee^{-\mathcal{S}(E_{mn})}|f(E_{mn},\omega_{mn})|^2$, within small frequency windows $\delta\omega \sim 0.2J$.

Aside from energy eigenstates, the ETH also describes the ergodic dynamics of non-equilibrium pure states seen in Fig.~\ref{fig:pure_thermo}. Using Eq.~\eqref{ETH-app}, the time average of an observable is found to be $\overline{ \braket{\hat{A}}} = \sum_n |\braket{E_n|\psi}|^2 A_{nn} =  A(\bar{E}) + \mathcal{O}(\Delta E^2/\bar{E}^2_*)$, while temporal fluctuations away from this value are proportional to the off-diagonal elements $A_{mn}$ and thus are exponentially suppressed~\cite{DAlessio2016}. The same applies to two-point correlation functions $C(t+\tau,t)$, which for large $t$ tend to their equilibrium, time-homogeneous value $C(t+\tau,t)\approx C(\tau)$. The latter is determined by the noise and response functions given in Eqs.~\eqref{noise_spectrum} and~\eqref{response_function} of the main text.
 
\subsubsection{Dynamical evolution}
\label{app:dynamics}

To compute dynamical quantities we solve the Schr\"odinger equation 
\begin{equation}
    \label{Schrodinger}
    \ii \partial_t \ket{\psi(t)} = \H \ket{\psi(t)},
\end{equation}
to obtain the state vector $\ket{\psi(t)}$, from which any observable $\braket{\hat{A}(t)} = \braket{\psi(t)|\hat{A}|\psi(t)}$ can be calculated. The decoherence function itself is obtained from the overlap $\varv(t) = \braket{\psi(t)|\psi'(t)}$, where the states $\ket{\psi(t)}$ and $\ket{\psi'(t)}$ are propagated under Eq.~\eqref{Schrodinger} with Hamiltonians $\hat{H}$ and $\hat{H}' = \hat{H}+g\hat{A}$, respectively. We integrate Eq.~\eqref{Schrodinger} using a standard fourth-order Runge-Kutta~\cite{Steinigeweg2015} algorithm with time step $J\delta t = 0.01$. This choice yields an excellent approximation to unitary evolution, e.g.~after an evolution time $Jt=50$ the normalisation $\braket{\psi(t)|\psi(t)}$ drops by less than 0.1\% at temperature $T=20J$ and by less than 1\% at $T=5J$. For very long evolutions, such as those required to compute $\varv(t)$ at weak coupling, we normalise the state after each time step to enforce unitarity. 

To compute two-point correlation functions, we follow the procedure described, for example, in Ref.~\cite{Steinigeweg2015}. For $t'>t$ we can write
\begin{align}
    \label{two_point_RK}
    \braket{\hat{A}(t')\hat{A}(t)} & = \braket{\psi(t_0)|\ee^{\ii\H(t'-t_0)} \hat{A}\ee^{-\ii\H(t'-t)}\hat{A}\ee^{-\ii\H(t-t_0)} |\psi(t_0)}\notag \\
    & = \braket{\psi(t')| \hat{A}|\phi(t')},
\end{align}
where $\ket{\psi(t_0)}$ and $t_0$ are the (arbitrary) initial state and time, and $\ket{\phi(t')} = \ee^{-\ii\H(t'-t)}\hat{A}\ket{\psi(t)}$ is obtained by propagating the Schr\"odinger equation~\eqref{Schrodinger} for a time interval $t'-t$ starting from the initial condition $\hat{A}\ket{\psi(t)}$. The case $t'<t$ is obtained by complex conjugation. 

\begin{figure}
    \centering
    \includegraphics[width=\linewidth]{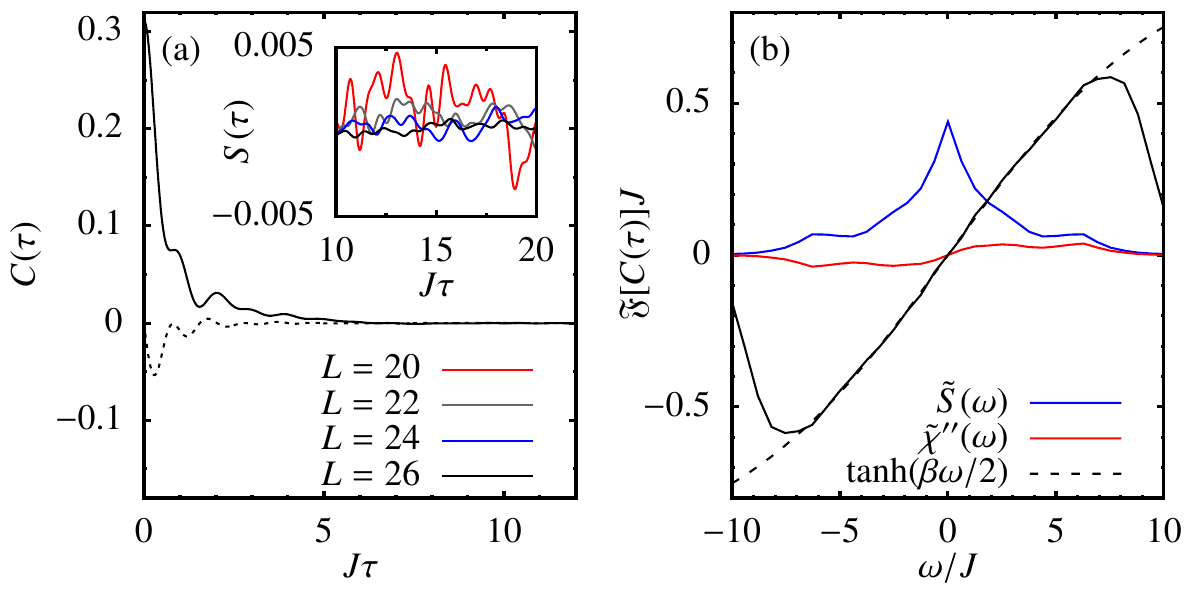}
    \caption{(a)~Real part (solid line) and imaginary part (dotted line) of the correlation function for a non-equilibrium pure state at energy $\bar{E}=-16J$, after thermalisation as described in Fig.~\ref{fig:pure_thermo} of the main text. The inset zooms in on the the real part at longer times for four different system sizes at the same temperature, $T=5J$. (b)~Noise and response functions for $L=26$, obtained from the correlation function by numerical Fourier transform up to a cutoff $J\tau_* = 10$. The black solid line shows their ratio, which approximately obeys the fluctuation-dissipation relation within the range of frequencies where $\tilde{\chi}''(\omega)$ is non-zero. The dashed line shows $\tanh(\beta \omega/2)$ for comparison, with $\beta(\bar{E}) = 0.20 J^{-1}$ obtained independently from a KPM calculation [see Sec.~\ref{app:KPM}]. \label{fig:SM_corr}}

\end{figure}

Fig.~\ref{fig:SM_corr}(a) shows an example of a correlation function computed in this way. Both real and imaginary parts of the correlation function decay over time, after which they execute small fluctuations near zero. These long-time fluctuations are a finite-size effect: they disappear rapidly with increasing system size [Fig.~\ref{fig:SM_corr}(a) inset]. To avoid this effect when constructing the noise and response functions by numerical Fourier transform, we sample the correlation functions only up to a time $\tau_*$, chosen to be $J\tau_*=10$. This is sufficiently large to capture all non-trivial dynamical features of $C(\tau)$, while generating a relatively smooth approximation to the noise and response function as shown in Fig.~\ref{fig:SM_corr}(b). Larger values of $\tau_*$ tend to generate spurious features in the frequency domain due to the long-time fluctuations of $C(\tau)$, which are most prevalent at lower temperatures. To extract a temperature from the Fourier data, as in Fig.~\ref{fig:pure_thermo}(f), we average the slope of $\tilde{\chi}''(\omega)/\tilde{S}(\omega)$ over the frequency range $|\omega|\leq 2J$, within which $\tanh(\beta\omega/2)\approx \beta\omega/2$ to an excellent approximation for the temperatures of interest.

\subsubsection{Microcanonical predictions}
\label{app:KPM}

Predictions of the microcanonical ensemble are evaluated using the kernel polynomial method (KPM)~\cite{Weisse2006,Yang2020}. The three quantities that we compute in this way are the density of states, $\Omega(E)$, the microcanonical expectation value, $A(E)$, and the local density of states, $|\psi(E)|^2$, given respectively by
\begin{align}
    \label{DOS}
   & \Omega(E) = \sum_n\delta(E-E_n),\\
   \label{A_mc}
   & A(E) = \frac{1}{\Omega(E)}\sum_n A_{nn} \delta(E-E_n),\\
   \label{LDOS}
   & |\psi(E)|^2 =  \sum_n |\braket{E_n|\psi}|^2 \delta(E-E_n).
\end{align}
The microcanonical entropy and temperature are extracted from the density of states via Boltzmann's relation $\mathcal{S}(E) = \ln W(E)$, where $W(E) = \Omega(E)\dd E$ corresponds to the number of microstates in a small energy interval $\dd E$.

The kernel polynomial method works by expanding the above functions in the basis of orthogonal Chebyshev polynomials, $T_n(E)$. Since these polynomials are defined only on the interval $E\in[-1,1]$, we first rescale the Hamiltonian spectrum to lie within this interval. Then, following the standard procedure detailed in Ref.~\cite{Weisse2006}, we approximate a continuous function of energy $\Xi(E)$ by evaluating a finite number of Chebyshev moments, $\mu_m = \int\dd E\, \Xi(E)T_m(E)$. The function is then reconstructed as the truncated basis expansion
\begin{equation}
    \Xi(E) \approx \frac{1}{\pi\sqrt{1-E^2}}\left[ g_0 \mu_0 +2 \sum_{m=1}^{M_{\rm Cheb}} g_m\mu_mT_m(E) \right],
\end{equation}
where $g_m$ are coefficients that decay with increasing $m$, which smooth the high-frequency oscillations (Gibbs phenomenon) that would otherwise result from truncating the expansion at finite order, $M_{\rm Cheb}$. We use the values of $g_m$ corresponding to the Jackson kernel, which is known to be optimal and effectively broadens the delta functions entering Eqs.~\eqref{DOS}--\eqref{LDOS} to Gaussians of width $\lesssim \pi/M_{\rm Cheb}$; see Ref.~\cite{Weisse2006} for details.

For example, the density of states is evaluated from the moments
\begin{equation}
\label{stochastic_trace}
    \mu_m = \Tr[T_m(\hat{H})] \approx \frac{1}{R}\sum_{r=1}^R \braket{r|T_m(\hat{H})|r}.
\end{equation}
The last step approximates the trace using a set of $R$ random vectors, where each component of the vector $\ket{r}$ is chosen independently from a Gaussian distribution. In a large Hilbert space only a few random vectors, say $R\lesssim 10$ for spin chain of length $L=26$, are typically needed to achieve excellent convergence within the relevant energy range for thermalisation. In the form~\eqref{stochastic_trace}, the moments can be computed recursively using the defining three-term relation of the Chebyshev polynomials, 
\begin{align}
    \label{Chebyshev_recursion}
 T_{m+1}(\hat{H}) = 2\hat{H}T_m(\hat{H}) - T_{m-1}(\hat{H}),
\end{align}
for $m\geq 1$, with $T_0(\hat{H}) = 1$ and $T_{1}(\hat{H}) = \hat{H}$. Acting this expression on a vector as $\ket{v_m}=T_m(\hat{H})\ket{r}$ generates the recursion relation $\ket{v_{m+1}} = 2\hat{H}\ket{v_m}-\ket{v_{m-1}}$. Each iteration thus requires a single matrix-vector multiplication, making the KPM a very efficient method for large, sparse matrices such as Eq.~\eqref{H_SF_app}.

\begin{figure}
    \centering
    \includegraphics[width=0.7\linewidth]{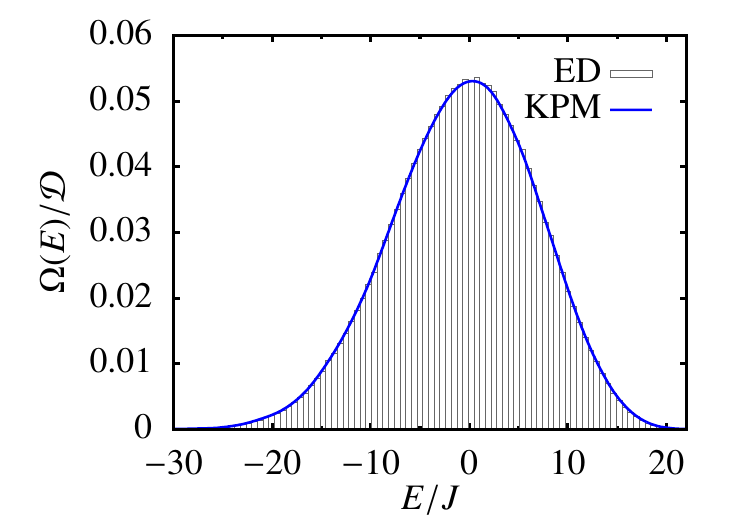}
    \caption{Density of states for a spin chain of $L=20$ sites. The bars show a normalised histogram of the energy levels obtained from exact diagonalisation. The blue solid line is the density of states obtained from KPM with $M_{\rm Cheb}= 100$ moments averaged over $R=150$ random vectors. Dividing $\Omega(E)$ by the Hilbert-space dimension ensures normalisation: $\int\dd E\, \Omega(E)/\mathcal{D} = 1$.}
    \label{fig:DOS}
\end{figure}

In all our calculations of the inverse temperature and microcanonical averages, we take $M_{\rm Cheb} = 100$ moments in total. This is sufficient to obtain an accurate, smooth approximation, as shown in Fig.~\ref{fig:DOS} for the density of states, for example. Larger values of $M_{\rm Cheb}$ can lead to spurious oscillations due to the underlying discreteness of the finite-dimensional Hilbert space, especially at low temperatures. For the local density of states shown in Fig.~\ref{fig:pure_thermo}(b), we take $M_{\rm Cheb}=250$ in order to better resolve the sharp features.

\subsection{Weak-coupling and long-time limit of the decoherence function}
\label{app:weak_coupling}

In this section we derive the weak-coupling expansion of the decoherence function, $\varv(t)$, and discuss its convergence to the long-time limit. To simplify the notation we set the initial time of the protocol to be $t_0=0$ in this section. Our starting point is the general expression for the decoherence function [c.f.~Eq.~\eqref{fidelity}]
\begin{equation}
    \label{dec_func}
    \varv(t) = \Braket{\ee^{\ii \H t}\ee^{-\ii(\H+g\hat{A})t}} = \Braket{{\rm T}\exp\left[-\ii g \int_0^t\dd t' \hat{A}(t')\right]},
\end{equation}
where we recognise the definition of the time-ordered exponential, with $\hat{A}(t) = \ee^{\ii\H t}\hat{A} \ee^{-\ii \H t}$ the Heisenberg-picture operator with respect to the Hamiltonian $\H$. Here, and in the following, expectation values are taken with respect to the initial state of the many-body system, $\braket{\bullet } = \Tr[\bullet \r]$, e.g.~$\r = \ket{\psi_0}\bra{\psi_0}$ in the case of a pure initial state as considered in the main text. Note, however, that our framework can be equally applied to any mixed state for which the one- and two-point functions of the operator $\hat{A}$ are approximately thermal.

We proceed by expanding Eq.~\eqref{dec_func} into time-ordered cumulants~\cite{Kubo1962} and neglecting terms of order $\mathcal{O}(g^3)$, which yields
\begin{align}
\label{cumulant_expansion}
    -\ln \varv(t) & \approx \ii g\int_0^t\dd t'\, \braket{\hat{A}(t')} + g^2\int_0^t\dd t'\int_0^{t'}\dd t''\, C(t',t'') \\
    & \approx \ii g t \overline{\braket{\hat{A}}} \notag \\ 
    & \quad + g^2\int_0^t\dd t'\int_0^{t'}\dd t''\, \left[S(t'-t'') + \chi''(t'-t'')\right],\notag\\
    &= \ii \Phi(t) + \frac{1}{2}\Gamma(t)
\end{align}
where the first line is the second-order cumulant expansion, while the second line follows by assuming the noise statistics are approximately stationary. On the final line, we moved to the Fourier domain, e.g.~$S(\tau) = \int\dd\omega \,\ee^{-\ii\omega \tau} \tilde{S}(\omega)/2\pi$, and defined
\begin{align}
     \label{Gamma_SM}
    & \Gamma(t) = 4g^2\int\frac{\dd \omega}{2\pi}\,\tilde{S}(\omega) \frac{\sin^2(\omega t/2)}{\omega^2},\\
   \label{Phi_SM}
   & \Phi(t)= g t\overline{\braket{\hat{A}}}  + g^2\int\frac{\dd \omega}{2\pi}\, \tilde{\chi}''(\omega) \frac{\sin(\omega t)-\omega t}{\omega^2}.
\end{align}

In order to analyse convergence to the long-time limit, it is convenient to remain in the time domain. Consider the second-order contribution to Eq.~\eqref{cumulant_expansion} in the stationary approximation
\begin{align}
    \label{second_cumulant_time}
   \int_0^t\dd t'\int_0^{t'}\dd t''\, C(t'-t'')  = \int_0^t\dd\tau\, (t-\tau)C(\tau),
\end{align}
where we introduced the variables $\tau = t'-t''$ and $\bar{t} = (t'+t'')/2$ and performed the trivial integral over $\bar{t}\in [\tau/2,t-\tau/2]$. Let $\tau_c$ denote the correlation time after which $C(\tau)$ has decayed to zero. Assuming that the improper integral $\int_0^\infty\dd\tau\, \tau C(\tau)$ exists, it gives a sub-leading (i.e.~constant) correction for large $t$. For $t\gg \tau_c$ we can therefore neglect this term and take the upper integration limit to infinity, obtaining 
\begin{equation}
    \label{second_cumulant_limit}
   \int_0^t\dd\tau\, (t-\tau)C(\tau)\approx  t\int_0^\infty \dd\tau\, C(\tau) = \frac{1}{2} \left[ \tilde{S}(0) -\ii \chi_{\hat{A}}\right],
\end{equation}
where $\chi_{\hat{A}} = \int\dd\omega\, \tilde{\chi}''(\omega)/\pi\omega$ is the thermodynamic susceptibility corresponding to the observable $\hat{A}$. This result also follows from taking $t\to\infty$ directly in Eqs.~\eqref{Gamma_SM} and \eqref{Phi_SM}. Eq.~\eqref{second_cumulant_limit} implies the emergence of pure exponential decay for times $t\gg \tau_c$, with the rate $\gamma = \lim_{t\to \infty} \dd\Gamma/\dd t = g^2 \tilde{S}(0)$. Therefore, self-consistency of the exponential approximation requires that the decoherence function evolves slowly so that $\gamma\tau_c\ll 1$. This condition is well satisfied by our examples, where the correlation function fully decays after a time $J\tau_c\lesssim 10$ [Fig.~\ref{fig:SM_corr}(a)], while the characteristic timescale for $\varv(t)$ is an order of magnitude longer [Fig.~\ref{fig:decoherence}(a)].

Note that the above arguments break down whenever $C(\tau)\sim \tau^{-p}$ with $p\leq 1$ for large $\tau$. In particular, diffusion in the thermodynamic limit implies that $S(\tau) = \Re[C(\tau)] \sim (D\tau)^{-d/2}$ in $d$ spatial dimensions with diffusion coefficient $D$, as discussed in Sec.~\ref{app:hydrodynamics}. For $d=1$, therefore, Eq.~\eqref{second_cumulant_time} is dominated by the second term in parentheses on the right-hand side, which implies the long-time behaviour $\Gamma(t)\sim t^{3/2}$. For $d=2$, we obtain the asymptotic scaling $\Gamma(t) \sim t\ln(Dt/\ell^2)$, where $\ell$ is the length scale characterising the probe [see Eq.~\eqref{diffusive_noise_in_time}]. In a finite system, however, diffusive dynamics persists up to the Thouless time, $t_{\rm T} \sim L^2/D$ ($L$ is the linear dimension of the system, defined in Sec.~\ref{app:hydrodynamics}), after which the correlation function $C(\tau)$ drops to zero (up to small fluctuations, see Fig.~\ref{fig:SM_corr}). One therefore expects Eq.~\eqref{second_cumulant_limit} to hold for low-dimensional diffusive systems after a correlation time scaling as $\tau_c \sim t_{\rm T}$. 

\subsection{Quantum Fisher information}
\label{app:Fisher}

In this section we discuss the quantum Fisher information and its contributions from the norm and phase of the decoherence function. Let us first briefly recap the meaning of the Fisher information in the context of parameter estimation. Suppose that the qubit probe is in the state $\hat{\rho}_{\rm q}(T)$, which depends on the temperature $T$. We consider a measurement described by a set of positive operators, $\{\hat{\Pi}(\xi)\}$, such that $\int\dd\xi\,  \hat{\Pi}(\xi) = 1$, where the possible measurement outcomes are labelled by the continuous index $\xi$ without loss of generality. A temperature estimate $T_{\rm est}(\bm{\xi})$ is constructed from the outcomes $\bm{\xi} = \{\xi_1,\ldots,\xi_M\}$ of a large number, $M$, of identical measurements on independent preparations of the state $\hat{\rho}_{\rm q}(T)$. We consider (asymptotically) unbiased estimators such as maximum likelihood estimation, which satisfy $\mathbb{E}[T_{\rm est}] = T$, where
\begin{equation}
    \label{expectation}
    \mathbb{E}[T_{\rm est}] = \int\dd\xi_1\cdots \int\dd\xi_M p(\xi_1|T) \cdots p(\xi_M|T) T_{\rm est}(\bm{\xi}),
\end{equation}
while $p(\xi|T) = \Tr[\hat{\Pi}(\xi) \hat{\rho}_{\rm q}(T)]$ denotes the probability of obtaining outcome $\xi$ in a single measurement. The expected error in the temperature estimate is thus $\Delta T^2 = \mathbb{E}[(T_{\rm est} - T)^2]$. This obeys the Cram\'er-Rao bound~\cite{Jaynes2003}, $\Delta T^2 \geq 1/M\mathcal{F}_T$, where the Fisher information is given by
\begin{equation}
    \label{Fisher_info}
    \mathcal{F}_T = \int\dd\xi \, p(\xi|T)\left(\frac{\partial \ln p(\xi|T) }{\partial T}\right)^2,
\end{equation}
which measures the sensitivity of the distribution to changes in the parameter $T$. The Fisher information depends on the choice of measurement basis, and is upper-bounded by the quantum Fisher information~\cite{Braunstein1994} (QFI), $\mathcal{F}_T\leq \mathcal{F}_T^Q$. The bound is saturated by the measurement of a specific observable: the symmetric logarithmic derivative (SLD), $\hat{\Lambda}_T$.

For a pure dephasing evolution, the qubit state is of the form~$\hat{\rho}_{\rm q} = \tfrac{1}{2}(1+\bm{v}\cdot \bm{\hat{\sigma}}),$ where $\bm{\sg} = (\sg_x,\sg_y,\sg_z)^{\sf T}$ is a vector of Pauli operators and $\bm{v} = (\Re[\varv],-\Im[\varv],0)^{\sf T}$ is the Bloch vector. Parameterising the decoherence function as $\varv = |\varv|\ee^{-\ii\phi}$, the QFI takes the form~\cite{Mitchison2020}
\begin{equation}
    \label{QFI_qubit}
    \mathcal{F}_T^Q  = \frac{1}{1-|\varv|^2}\left(\frac{\partial |\varv|}{\partial T}\right)^2 + |\varv|^2 \left(\frac{\partial \phi}{\partial T}\right)^2  = \mathcal{F}^\parallel_T + \mathcal{F}^\perp_T .
\end{equation}
These two terms respectively correspond to the Fisher information for measurements of $\sg_\parallel = \cos(\phi) \sg_x - \sin(\phi) \sg_y$ and $\sg_\perp = \cos(\phi)\sg_y + \sin(\phi)\sg_x$, i.e.~the bases parallel and perpendicular to $\bm{v}$ in the equatorial plane of the Bloch sphere [see Fig.~1 of the main text]. Up to irrelevant additive and multiplicative factors, the SLD is given by 
\begin{align}
    \label{SLD}
    & \hat{\Lambda}_T \propto \cos (\varphi)\sg_\parallel + \sin (\varphi) \sg_\perp, \\ 
    & \tan (\varphi) =\frac{ |\varv|(1-|\varv|)^2 \partial_T \phi}{\partial_T|\varv|} \notag .
\end{align}

\begin{figure}
    \centering
    \includegraphics[width=\linewidth]{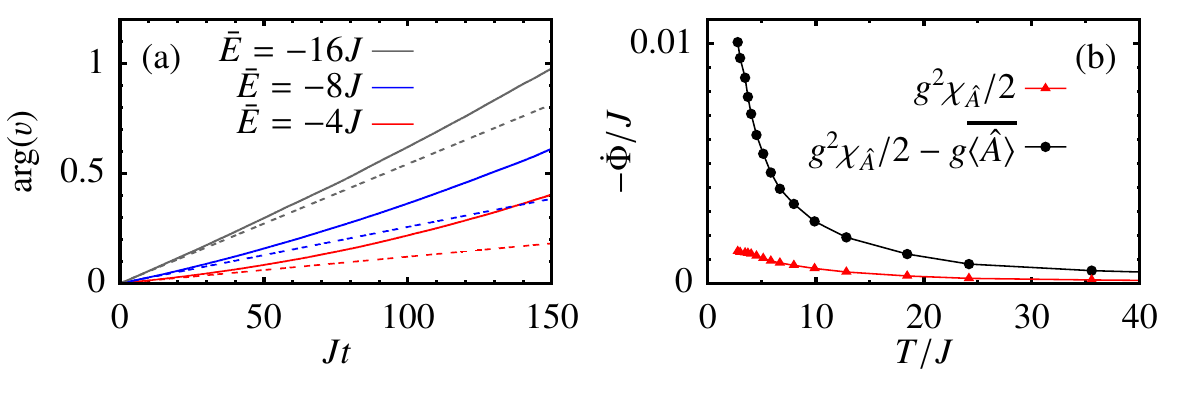}
    \caption{(a) Phase of the decoherence function with coupling strength $g=0.2J$ at three different temperatures, as in Fig.~\ref{fig:decoherence}(a) of the main text. Solid lines show the exact calculation and dashed lines show the corresponding asymptotic approximation, $\phi = \Phi(t)\to (g\overline{\braket{\hat{A}}}-g^2\chi_{\hat{A}}/2)t$ [see Eq.~\eqref{Phi_SM}], obtained from a dynamical calculation of $C(\tau)$ [see Sec.~\ref{app:dynamics}] (b) Temperature dependence of the asymptotic phase accumulation rate, $\dot{\Phi} = \dd\Phi/\dd t$, in the weak-coupling limit, as extracted from a dynamical calculation of $\chi''(\tau)$.}
    \label{fig:SM_phase}
\end{figure}

The main text considers thermometry using measurements of the norm $|\varv(t)|$, neglecting the phase $\phi$. In Fig.~\ref{fig:SM_phase}(a) we plot the phase at the same coupling $g=0.2J$ as considered in the main text. The results are compared to the asymptotic weak-coupling approximation, $\phi\approx \Phi(t) \to (g \overline{\braket{\hat{A}}} - g^2\chi_{\hat{A}}/2)t$ in Fig.~\ref{fig:SM_phase}. The exact results noticeably diverge from the weak-coupling prediction at longer times, presumably because of higher-order effects not captured by the perturbative expansion to order $\mathcal{O}(g^2)$. The difficulty of predicting the phase accurately with perturbative approximations is well known, e.g.~from studies of the Fermi edge singularity where the cumulant expansion for $\phi$ diverges already at second order~\cite{Mitchison2020}. In Fig.~\ref{fig:decoherence}(b) we plot the temperature dependence of phase contributions in the weak-coupling limit, showing that they depend more weakly on temperature than the corresponding decoherence rate in this example. As a consequence, the QFI~\eqref{QFI_qubit} is dominated by the first contribution, $\mathcal{F}^\parallel_T$, as shown in Fig.~\ref{fig:decoherence}(d) of the main text. The corresponding measurement of $\sg_\parallel$ could be enacted by tuning the phase of the final $\pi/2$-pulse, $\theta$, to equal the phase of the decoherence function, i.e. $\theta=\phi$, as discussed in the main text. 

To obtain a smooth prediction for the Fisher information in the weak-coupling limit, as shown in Fig.~\ref{fig:decoherence}(d), we construct a spline interpolation of the data shown in Figs.~4(c) and~\ref{fig:SM_phase}(b). The derivatives in Eq.~\eqref{QFI_qubit} are then approximated by a first-order finite difference with $\delta T\approx 0.2J$.

\subsection{Hydrodynamic response function}
\label{app:hydrodynamics}

In this section, we discuss the form of the response function that follows from diffusive hydrodynamics, and the resulting temperature dependence of the decoherence rate for large systems. We closely follow the classic analysis of Kadanoff \& Martin~\cite{Kadanoff1963,Forster2018}, generalising it to a large but finite system in $d$ spatial dimensions. We consider a translation-invariant system of volume $L^d$ under periodic boundary conditions, described by the Hamiltonian $\hat{H}$. Let $\hat{A} = \int\dd\rr\, u(\rr) \hat{n}(\rr)$ be a local observable written in terms of the density, $\hat{n}(\rr)$, of a globally conserved charge, $\hat{N}=\int\dd\rr\, \hat{n}(\rr)$, such that $[\hat{H},\hat{N}]=0$. The slowly varying function $u(\rr)$ represents the spatial profile of the probe, which we normalise as $\int\dd\rr\, u(\rr) = 1$ without loss of generality. 

Translation invariance allows us to separately analyse each Fourier component of the density, $\hat{n}_\kk = \int\dd\rr \, \ee^{-\ii\kk\cdot\rr} \hat{n}(\rr)$, where the discrete wavevector $\kk$ has components $k_j = 2\pi n_j/L$ with $n_j\in \mathbb{Z}$ and $j=1,\ldots d$. For any state that is invariant under space and time translations we can then define the density response function via
\begin{equation}
    \label{density_response_function}
    \frac{1}{2L^d}\braket{[\hat{n}_\kk(t+\tau),\hat{n}_{-\kk'}(t)]}= \delta_{\kk\kk'} \chi''_\kk(\tau),
\end{equation}
with $\tilde{\chi}''_\kk(\omega)$ the Fourier transform of $\chi''_\kk(\tau)$. Note that $\chi''_{\kk =0}(\tau)=0$ identically as a consequence of the conservation of total charge, $\hat{N} = \hat{n}_{\kk=0}$. The probe observable reads $\hat{A} = L^{-d}\sum_\kk c_{-\kk} \hat{n}_\kk$, with the corresponding response function $\tilde{\chi}''(\omega) = L^{-d}\sum_{\kk\neq 0} |u_\kk|^2 \tilde{\chi}_\kk''(\omega)$, where $u_\kk$ is the Fourier transform of $u(\rr)$. For example, if $u(\rr) \propto e^{-r^2/2\ell^2}$ is a Gaussian of width $\ell$, $u_\kk = \ee^{-\ell^2k^2/2}$ cuts off wavevectors $k\gg \ell^{-1}$; other smooth profiles show similar behaviour. Therefore, so long as $u(\rr)$ is slowly varying (i.e. $\ell$ is large), only long-wavelength diffusive modes contribute significantly to $\tilde{\chi}''(\omega)$. 

To find the response function associated with diffusion, a small density modulation is introduced by adiabatically switching on a weak force, $F(\rr)$, and then suddenly removing it at $t=0$. This is modelled by the potential
\begin{equation}
    \label{adiabatic_perturbation}
    \hat{V}(t) = - \Theta(-t)\ee^{\varepsilon t} \int\dd \rr\, F(\rr) \hat{n}(\rr),
\end{equation}
where $\Theta(t)$ is the unit step function and $\varepsilon\to 0$ is a small, non-negative convergence parameter. The resulting density deviation is then computed from linear-response theory, starting from an equilibrium state at the initial time $t_0\to-\infty$. For $t<0$, the system adiabatically follows the slow perturbation, thus remaining in thermal equilibrium, $\langle \hat{n}_\kk(t\leq 0)\rangle=\langle \hat{n}_\kk\rangle_{\rm eq}$. Meanwhile, linear-response theory yields $\langle \hat{n}_\kk(t\leq 0)\rangle = \chi_\kk F_\kk$, where
\begin{equation}
    \label{thermodynamic_susceptibility}
    \chi_\kk = \int\dd\omega \, \frac{\tilde{\chi}''_\kk(\omega)}{\pi\omega}  = \left. \frac{\partial  \langle \hat{n}_\kk\rangle_{\rm eq}}{\partial F_{\kk}}\right\rvert_{F=0},
\end{equation}
which is identified as the thermodynamic susceptibility. For $t>0$, we obtain
\begin{equation}
\label{}
    \braket{\hat{n}_\kk(t)} = \int\dd \omega\,\frac{\tilde{\chi}''_\kk(\omega) F_\kk }{\pi \omega} \ee^{-\ii\omega t},
\end{equation}
which, assuming a slowly varying $F(\rr)$, should evolve according to the diffusion equation $(\partial_t + Dk^2)\braket{\hat{n}_\kk(t)}=0$. This is readily solved by $\braket{\hat{n}_\kk(t)} = \chi_\kk F_\kk \ee^{-Dk^2 t}$, given the initial condition at $t=0$. Comparing these two solutions for $\braket{\hat{n}_\kk(t)}$ in the Laplace domain, we get
\begin{equation}
    \int\frac{\dd\omega'}{\ii\pi} \frac{\tilde{\chi}''_\kk(\omega')}{\omega'(\omega'-z)} = \frac{\chi_\kk}{Dk^2-\ii z},
\end{equation}
where the Laplace variable obeys $\Im z>0$. Finally, by continuing $z\to \omega+\ii 0$ to the real axis one deduces the density response function $\tilde{\chi}''_\kk(\omega)$. The response function for $\hat{A}$ then follows as
\begin{equation}
    \label{diffusive_response}
    \tilde{\chi}''(\omega) = \frac{1}{L^d}\sum_{\kk\neq 0} \frac{\chi_\kk |u_\kk|^2 Dk^2\omega}{\omega^2 + (Dk^2)^2}.
\end{equation}

In a sufficiently large system, the summation is well approximated by an integral using the standard prescription $L^{-d} \sum_\kk  \to (2\pi)^{-d}\int\dd \kk$. If we also assume that $u_\kk$ samples only small wavevectors such that $\chi_\kk \approx \lim_{\kk\to 0}\chi_\kk \equiv \chi_0$, then the integral can be carried out explicitly. At small positive frequencies, we obtain the limiting behaviour
\begin{equation}
    \label{spectral_density_dimensions}
   \tilde{ \chi}''(\omega) \sim \begin{cases}
    \sqrt{\omega} & (d=1)\\ 
    -\omega\ln(\ell^2\omega/D)  & (d=2)\\
    \omega & (d\geq 3).
    \end{cases}
\end{equation}
For $d=2$ only, the low-frequency response function depends sensitively on the ultraviolet (UV) cutoff scale, $D/\ell^2$, where $\ell$ is the width of the probe function $u(\rr)$ discussed above. For $d\geq 3$, the response is Ohmic and the zero-frequency limit of the symmetrised noise $\tilde{S}(\omega) = \coth(\beta\omega/2)\tilde{\chi}''(\omega)$ is well behaved. The long-time decoherence rate thus follows from the limit $\gamma = \lim_{\omega\to 0} 2\tilde{\chi}''(\omega)/\beta\omega$, which can be taken directly in Eq.~\eqref{diffusive_response} to obtain
\begin{equation}
    \label{gamma_3D}
    \gamma =\frac{2g^2 T}{D} \int\frac{\dd\kk}{(2\pi)^3}  \frac{\chi_\kk |u_\kk|^2}{k^2} \approx \frac{2\bar{g}^2 T \chi_0}{D},
\end{equation}
where the final approximation follows from the assumption that $u_\kk$ samples only long wavelengths, and we defined the renormalised coupling $\bar{g}^2 = g^2 \int\dd\kk |u_\kk|^2/(8\pi^3 k^2),$ which depends only on properties of the probe.

For $d<3$, the low-frequency limit of Eq.~\eqref{diffusive_response} is ill-behaved in an infinite system. For any finite system, however, the lower $k$-integration limit is cut off by the exclusion of $\kk=0$. In 1D this yields
\begin{equation}
    \label{gamma_1d}
    \gamma = \frac{2g^2T}{\pi D}\int_{2\pi/L}^\infty \dd k \, \frac{\chi_k|u_k|^2}{k^2} \sim L,
\end{equation}
since for large $L$ the integral is dominated by its lower limit. A similar argument in 2D yields $\gamma \sim \ln(L/\ell)$, where again the UV cutoff appears explicitly for dimensional reasons. This diverging zero-frequency noise in low dimensions originates from the long diffusive tail of the correlation function in time. To see this explicitly, we take the Fourier transform of Eq.~\eqref{diffusive_response} to obtain, for $\tau>0$,
\begin{align}
    \label{diffusive_response_in_time}
    \ii\chi''(\tau) & = \frac{1}{2L^d}\sum_{\kk\neq 0} \chi_\kk |u_\kk|^2 Dk^2 \ee^{-Dk^2\tau}
\end{align}
Assuming as above that $\chi_
\kk|u_\kk|^2\approx \chi_{0}\ee^{-\ell^2k^2/2}$, we obtain $\chi(\tau)\propto (\ell^2+2D\tau)^{-(d/2+1)}$ in the thermodynamic limit. The leading-order contribution to $C(\tau)$ arises from the corresponding symmetrised noise, which for $\tau\to\infty$ can be approximated from the low-frequency response as
\begin{align}
    \label{diffusive_noise_in_time}
    S(\tau) & = \int \frac{\dd\omega}{2\pi} \coth(\beta\omega/2) \tilde{\chi}''(\omega) \ee^{-\ii\omega \tau} \notag \\
    & \approx  \int \frac{\dd\omega}{2\pi} \frac{2 \tilde{\chi}''(\omega)}{\beta\omega} \ee^{-\ii\omega \tau} \notag \\
   & = \frac{1}{\beta L^d}\sum_{\kk\neq 0} \chi_\kk |u_\kk|^2 \ee^{-Dk^2\tau},
\end{align}
from which we deduce $S(\tau) \sim  \left(\ell^2 + 2D\tau \right)^{-d/2}.$

\end{document}